\documentclass{ws-procs975x65}

\def\be{\begin{equation}}
\def\ee{\end{equation}}
\def\bea{\begin{eqnarray}}
\def\eea{\end{eqnnarray}}

\def\e{\epsilon}
\def\k{\kappa}
\def\s{\sigma}
\def\z{\zeta}
\def\Floc{f_{NL}^{loc}}

\begin{document}

\title{MULTIFIELD COSMOLOGICAL PERTURBATIONS AT THIRD ORDER AND THE EKPYROTIC TRISPECTRUM}

\author{SEBASTIEN RENAUX-PETEL}

\address{APC (UMR 7164, CNRS, Universit\'e Paris 7)\\
Paris, 75205 Paris Cedex 13, France\\
E-mail: renaux@apc.univ-paris7.fr}

\begin{abstract}
We explain the motivation and main results of our work in Ref.~\refcite{Lehners:2009ja}. Using the covariant formalism, we derive the equations of motion for adiabatic and entropy perturbations at third order
in perturbation theory for cosmological models involving two
scalar fields, and use these equations to calculate the
trispectrum of ekpyrotic and cyclic models. The non-linearity
parameters $f_{NL}$ and $g_{NL}$ are found to combine to leave a very distinct
observational imprint.
\end{abstract}


\bodymatter

\section{Introduction}

The analysis of non-Gaussianity \cite{Komatsu:2009kd}, {\it i.e.} the deviation from perfect
Gaussian statistics, is becoming a strong
discriminator of competing models of the early universe. Whilst most efforts until now have been devoted to the study of the bispectrum -- the Fourier transform of the 3-point function of the curvature perturbation $\zeta$ -- recent progress in observational cosmology is such that it may be possible in future to detect the trispectrum (4-point correlation function of $\zeta$), if it is large enough. However, the calculational technology required to make the corresponding predictions is not yet in place. In particular, analytic formulae can be obtained from the $\delta N$ formalism, the most widely used technique, only in very special cases and furthermore its numerical implementation is demanding. Here we report on a recent development: the derivation of the equations of motion to third order in perturbation theory, which are needed to calculate the non-Gaussianity in two-scalar-field cosmological models. The resulting equations are generally applicable and well suited to numerical analyses, as demonstrated by our calculation of the trispectrum generated by ekpyrotic and cyclic cosmological models --- scenarios in which the primordial fluctuations are created during a collapsing phase before the Big-Bang (see Ref.~\refcite{Lehners:2008vx} for a review).

\section{Multifield cosmological perturbations}

We consider cosmological models involving two fields with standard kinetic terms. In such models large non-Gaussianities can be generated because of the classical nonlinear evolution outside the horizon: in general multiple field models, the scalar perturbations can be decomposed into (instantaneous) adiabatic and entropy modes by projecting, respectively, parallel and perpendicular to the background trajectory in field space \cite{Gordon:2000hv}. As the entropy fields are not submitted to the slow-roll requirements, they need not be almost free fields and can develop large nonlinearities. If they are light enough to be quantum mechanically excited during inflation, they develop super-Hubble fluctuations that can be transferred, together with their nonlinearities, to the adiabatic mode or curvature perturbation, if there is a turn in field space (or even more generically with non standard kinetic terms \cite{Langlois:2008mn}). The conversion from entropy to curvature perturbations can also itself be nonlinear. This results in possibly large non-Gaussianities of the so-called local form, characterized in the simplest models by two non-linearity parameters $\Floc$ and $g_{NL}$ via an expansion of the curvature perturbation $\zeta$ in terms of its linear, gaussian part $\zeta_L,$
\be 
\z = \z_L + \frac{3}{5} \Floc \z_L^2 + \frac{9}{25} g_{NL} \z_L^3. 
\ee
The determination of $\Floc$ (resp. $g_{NL}$) requires the derivation of the coupled equations of motion for adiabatic and entropy perturbations on large scales to second order (resp. third order) in perturbation theory. This rather intricate task can be fulfilled by means of the covariant formalism, in which one  easily derives exact fully nonlinear equations for covectors that can then be expanded into perturbation theory. Langlois and Vernizzi used this method to derive the relevant equations to second order in Ref.~\refcite{Langlois:2006vv} (see also Ref.~\refcite{RenauxPetel:2008gi}), and their results were subsequently applied to the determination of the bispectrum ($\Floc$) in ekpyrotic and cyclic models in Refs.~\refcite{Lehners:2007wc,Lehners:2008my}. In Ref.~\refcite{Lehners:2009ja}, Jean-Luc Lehners and I have extended these two works to third order in perturbation theory. It is worth emphasizing that our analysis is quite general: we did not specify an underlying
potential and do not use approximations other than the large
scale limit. In particular, although we have applied the formalism to a
particular scenario of a contracting universe, it is equally
well suited to the study of cosmological perturbations
generated during a period of inflation.

\section{The ekpyrotic trispectrum}

In ekpyrotic scenarios, the curvature perturbation is generated through the entropic mechanism explained above.
Broadly speaking, there are two limiting cases that are of special
interest: the first is where the bending of the trajectory occurs after the
ekpyrotic phase, during the approach to the big crunch. In this
case, the kinetic
energy of the scalar fields is the dominant contribution to the
total energy density while the conversion takes place: we call this case ``kinetic conversion''. The other limiting case is where the conversion occurs during
the ekpyrotic phase --- ``ekpyrotic conversion''. During the ekpyrotic phase, we adopt the following parametrization of the
potential: 
\be 
V_{ek}=-V_0
e^{\sqrt{2\e}\s}[1 +\e s^2 +\frac{\k_3}{3!}\e^{3/2} s^3
+\frac{\k_4}{4!}\e^2 s^4+\cdots],
\ee
where $\s$ denotes the adiabatic direction, $s$ denotes the ``entropy'' direction, we expect $\k_3,\k_4
\sim {\cal O}(1)$ and where $\e \sim {\cal O}(10^{2})$ is
related to the ekpyrotic equation of state $w_{ek}$ via
$\e=3(1+w_{ek})/2.$ 

Then, for {\em kinetic conversions} lasting of the
order of one e-fold of contraction of the scale factor, we find
the following approximate fitting formula (for which there is an analytical understanding \cite{Lehners:2009qu}) for the
third order non-linearity parameter $g_{NL}$ (and we include $\Floc$ for completeness \cite{Lehners:2007wc,Lehners:2008my})
: 
\begin{eqnarray} 
 \Floc &\sim& \frac{3}{2} \, \k_3 \sqrt{\e} +5 \\
g_{NL} &\sim& 
\e \left( \frac{5}{3} \k_4+\frac{5}{4}  \k_3^2- 40\right). 
\end{eqnarray} 
 Note that when $\k_3$ and $\k_4$ are small, $g_{NL}$ is always
negative. Hence, even though $\Floc$ is small in that case,
$g_{NL}$ is negative and typically of order a few thousand, so
that any accidental degeneracy at the level of $\Floc$ between
simple inflationary and cyclic models is very likely to be
broken at the level of the trispectrum. More generally, unless
$|\Floc|$ turns out to be quite large, one would typically
expect $g_{NL}$ to be negative, as obtaining a positive
$g_{NL}$ in that case would require unnaturally large values of
$\k_4.$

For {\em ekpyrotic conversions}, the $\delta N$ formalism is easily applicable and gives approximate formulae for the
non-linearity parameters that we were able to confirm to a good accuracy by numerically solving the equations of motion up to third order. In that case $\Floc$ is always large and negative while $g_{NL}$ is always
positive. The current limits on $\Floc$ \cite{Smith:2009jr} are therefore already putting some
strain on this particular mode of conversion.

\section{Conclusion}

Motivated by huge observational perspectives in coming years concerning non-Gaussianity measurements, we have developed in Ref.~\refcite{Lehners:2009ja} the necessary tools to compute the trispectrum generated in  two-field cosmological models. Our results are generally applicable and well suited for numerical analyses and we used them to calculate the trispectrum generated in ekpyrotic and cyclic models. The combined consideration of the bi- and trispectrum in that case results in a distinct observational imprint that should enable one to select or rule these models on observational grounds in the foreseeable future.

\end{document}